\documentclass[a4paper,fleqn]{cas-dc}

\usepackage[numbers]{natbib}
\usepackage{amsmath}
\usepackage{booktabs}
\usepackage{multirow}
\usepackage{xcolor}
\usepackage{xurl}
\usepackage{orcidlink}
\usepackage{soul}

\newcommand{\jev}{\textsc{Jev}}
\newcommand{\jevids}{\textsc{Jev-IDS}}

\begin{document}

\let\WriteBookmarks\relax
\def\floatpagepagefraction{1}
\def\textpagefraction{.001}

\shorttitle{Jev-IDS: System One Models for Network Intrusion Detection}
\shortauthors{S. E. Quincozes}

\title[mode=title]{Jev-IDS: A System One Model for Network Intrusion Detection in Label-Scarce and Zero-Day Scenarios}

\author[1]{Paulo Severo\orcidlink{0000-0003-4945-3329}}
\author[2]{Silvio E. Quincozes\orcidlink{0000-0001-6793-4033}}\cormark[1]
\ead{silvioquincozes@unipampa.edu.br}
\author[3]{Amanda Dias\orcidlink{0000-0001-6793-4033}}

\affiliation[1]{organization={Graduate Program in Software Engineering, Federal University of Pampa (UNIPAMPA)},
                 city={Alegrete},
                 state={Rio Grande do Sul},
                 country={Brazil}}

\cortext[cor1]{Corresponding author.}

\begin{abstract}
Machine-learning Network Intrusion Detection Systems (IDS) depend on substantial labeled datasets and task-specific training, whereas Large Language Models (LLMs) detection can analyze flow records directly but incurs higher inference cost and latency, with less constrained outputs. This paper presents \jevids{}, an open experimental general NIDS based on the Jev System One Model (SOM) to detect zero day intrusions Under label scarcity. \jevids{} serializes one flow per request and asks \jev{} two questions: a binary attack probability and a finite-choice traffic category. Our results show that, at $k=1$, \jev{} was 4.8 times faster and 3.8 times cheaper than GPT-5.6 Luna, with 1.5 times higher novel-attack recall; it also produced 15 times fewer false alarms than a low-data Random Forest. Across 5,400 decisions on a 300-flow NSL-KDD pilot split, \jev{} achieved $F_1=0.859$, precision 0.941, recall 0.790, and novel-attack recall 0.838. Increasing $k$ to 2 reduced its $F_1$ to 0.839. 
\end{abstract}

\begin{highlights}
\item Jev-IDS reproducibly evaluates a System One Model as a flow-based NIDS.
\item The protocol jointly measures detection, labeled-data use, latency, and cost.
\item Frozen splits distinguish known attacks from attacks absent from the pool.
\item A 5,400-decision pilot compares JEV, GPT-5.6 Luna, and Random Forest.
\end{highlights}


\maketitle

\section{Introduction}
\label{sec:introduction}

Network Intrusion Detection Systems (NIDSs) monitor network traffic to identify malicious activity and support timely incident response. Learning-based NIDSs use traffic features to recognize attack patterns, offering an alternative to manually specifying every detection rule. However, their practical value depends on more than classification accuracy: missed attacks leave services exposed, false alarms increase the workload of security teams, and inference overhead limits the volume of traffic that can be inspected \cite{sommer2010closed}.

Conventional supervised detectors rely on labeled data that adequately represent the traffic and attacks they will encounter. Obtaining such data is particularly difficult when attacks are rare or newly observed, motivating approaches that can adapt from only a few examples \cite{khan2026metalearning}. Recent advances in Large Language Models (LLMs) provide an opportunity to reduce this dependence on task-specific training. By interpreting text-encoded flows, instructions, and optional demonstrations, prompted LLMs can perform intrusion classification without updating their model parameters \cite{zhang2024llmnids,houssel2024explainable}.

Despite this flexibility, LLM-based detection introduces limitations that affect its operational usefulness. Comparative evidence shows that LLMs can incur substantially higher inference overhead without outperforming strong conventional classifiers \cite{mehavilla2026evaluating}. Their generated responses also require output control before they can be reliably consumed by monitoring software. Grammar-constrained approaches already produce structured verdicts and attack scores \cite{rehman2025flows}, but enforcing an output format does not establish that the underlying classification is correct or computationally efficient. These limitations leave a practical challenge: detecting attacks with scarce labeled evidence while maintaining reliable decisions, low latency, and affordable inference. 

More recently, System One Model (SOM) was introduced as a model family designed for typed probabilistic decisions instead of free-form text generation. \jev{} is the first model from this new class, developed by TypeSafe AI and launched in September 2026~\cite{typesafe2026jev}. Their interface is potentially well matched to intrusion detection: the input is a structured program state, the possible outputs are specified in advance, and the model returns probabilities that can drive an alerting or escalation policy. Nevertheless, the performance claims currently available for this model family concern general automation workflows rather than NIDS. It remains unknown whether such a model can distinguish benign and malicious flows, whether a small number of labeled examples improves its decisions, and whether its operational advantages persist in a controlled NIDS comparison.

This paper introduces \jevids{}, an open experimental NIDS built around \jev{} to study intrusion detection with scarce labels and previously unseen attack categories. For each sample, the system requests an attack probability and a traffic category from a predefined set, without task-specific fine-tuning. Our objective is to determine whether this approach improves the detection--latency--cost trade-off under label scarcity and for attacks absent from the example pool. We compare \jev{} with a structured-output GPT-5.6 Luna detector and a low-data Random Forest using the same flow attributes, labeled examples, and binary decision threshold.

Our main contributions are summarized as follows:
\begin{itemize}
    \item We propose \jevids{}, the first flow-based NIDS that integrates a SOM through typed probabilistic and finite-choice questions (Section~\ref{sec:architecture});
    \item We define a controlled evaluation protocol for varying the number of labeled examples and assessing simulated zero-day scenarios (Section~\ref{sec:method});
    \item We compare three detectors on two intrusion datasets using one or two labeled examples per category and three seeds, assessing detection errors, latency, and estimated cost (Section~\ref{sec:results});
    \item We release the implementation, prompts, frozen splits, and per-flow pilot predictions to support reconstruction of the reported comparisons \cite{jevids2026software} (Section~\ref{sec:method}).
\end{itemize}

The remainder of this paper is organized as follows. Sections~\ref{sec:background} and \ref{sec:related-work} presents the background and related work. Section~\ref{sec:architecture} presents \jevids{}, and Section~\ref{sec:method} describes the experimental methodology. Section~\ref{sec:results} reports the results, followed by their discussion in Section~\ref{sec:discussion}. Finally, Section~\ref{sec:conclusion} concludes the paper.

\section{Background}
\label{sec:background}
This section introduces the concepts required to frame the evaluation, focusing on flow classification under limited supervision, LLM-based decision components, and SOMs. These concepts provide the basis for contrasting autoregressive generation with typed probabilistic decision interfaces for network intrusion detection.

\subsection{Flow classification with limited supervision}

Flow-based NIDSs represent network connections through attributes such as protocol, duration, byte counts, and packet statistics. Conventional supervised detectors learn a mapping from these attributes to benign traffic, malicious traffic, or specific attack categories by updating their parameters on labeled examples. Their effectiveness therefore depends on whether the training data adequately represent the traffic patterns encountered during deployment~\cite{mehavilla2026evaluating}.

Under label scarcity, in-context inference offers a different adaptation mechanism. Labeled examples are supplied with each request instead of being used to update the model parameters. This difference also affects evaluation. If an attack type appears in the example pool, classifying a new flow from that type measures few-shot generalization. If the attack type is absent, the experiment measures the ability to recognize an unseen attack~\cite{khan2026metalearning}. 

This work evaluates both settings: few-shot classification for represented attack types and novel-attack detection for attack types absent from the example pool.

\subsection{Language models as decision components}

Prompting allows a pretrained LLM to interpret a task description and classify a serialized record without task-specific parameter updates. Zero-shot inference supplies instructions alone~\cite{wei2022finetuned}. Few-shot inference also includes labeled demonstrations in the prompt~\cite{brown2020language}. In agent architectures, the same mechanism supports tool selection, information retrieval, and decisions about subsequent actions. The surrounding application maintains the state and executes the selected operations~\cite{chowa2026language}. This separation between model judgments and program execution provides a conceptual basis for using general-purpose models as components of a detection pipeline.

Autoregressive LLMs express a decision by generating tokens sequentially. A schema or decoding grammar can restrict this sequence to valid labels and numerical fields, making the response easier to process programmatically~\cite{rehman2025flows}. However, these constraints only control the response format; they do not remove the token-generation process. Even a structured response may contain field names, delimiters, explanations, or other tokens that are unnecessary when the IDS only needs a class label and an attack probability.

These additional tokens are not operationally free. Autoregressive decoding generates them sequentially, increasing inference time and computational demand. This overhead is especially relevant for high-rate NIDSs, as LLM-based detectors process substantially fewer flows per second than conventional ML models~\cite{mehavilla2026evaluating}. Commercial APIs also charge per token, meaning that unnecessary output tokens contribute to the cost of each request. This cost accumulates as the number of processed flows increases~\cite{guan2026notrain}. Therefore, output validity, predictive correctness, latency, resource consumption, and monetary cost must be evaluated separately.

\subsection{System One models and typed decisions}

TypeSafe AI introduced the term SOM for models designed to return bounded, probabilistic judgments directly, with \jev{} as its first publicly accessible implementation~\cite{typesafe2026jev}. The documented interface receives a state and a set of typed questions. A \texttt{noul} question returns a value in $[0,1]$ for a binary proposition, while a \texttt{choice} question returns an option and probabilities over a predefined set. Multiple questions can share the same state and are evaluated independently. The application combines their answers through ordinary program logic~\cite{typesafe2026primitives}.

This interface separates semantic judgment from response generation: an application defines the admissible answers before inference and receives values suitable for thresholding, ranking, or routing. It also separates the decision model from a complete agent, which additionally requires state management and an execution policy. For intrusion detection, the relevant primitives are a binary maliciousness judgment and a choice among traffic categories; the mapping from these outputs to alerts remains an application-level decision.

Recent applications illustrate this design outside network security. Jev-Mem uses \jev{} to organize agent memories and control retrieval, while an LLM synthesizes the final answer~\cite{jiang2026jevmem}. Another study formulates information extraction from police crash narratives as typed questions with predefined response options~\cite{rafe2026calibrated}. Both are September 2026 preprints and provide examples of SOM use in memory management and document analysis. They establish application contexts for the interface rather than evidence of its effectiveness for network intrusion detection.

Finally, a typed probability is not automatically a calibrated probability. Calibration concerns agreement between predicted probabilities and observed outcome frequencies. The crash-narrative study finds that calibration must be assessed for each model and task, even when probabilities are returned directly~\cite{rafe2026calibrated}. Thus, a constrained answer space simplifies integration, while detection accuracy and probability reliability still require empirical evaluation.

\section{Related work}
\label{sec:related-work}

This section summarizes the studies closest to \jevids{} in terms of inference-time adaptation, direct traffic classification, or the use of an LLM within an operational NIDS. It states the role of the language model explicitly because an LLM that classifies each flow has different latency and cost implications from one that explains alerts or occasionally advises model adaptation. Reported detection scores are omitted because the studies use different datasets, class distributions, supervision budgets, and evaluation splits.

\subsection{Language models as traffic classifiers}

Early studies on IDSs using LLMs established that network records can be serialized as text and classified through prompting. Zhang et al.~\cite{zhang2024llmnids} examined in-context examples for automatic intrusion detection, while Houssel et al.~\cite{houssel2024explainable} evaluated zero-shot and fine-tuned models and also studied generated explanations. Bui et al.~\cite{bui2024systematic} subsequently compared prompting, retrieval, and fine-tuning on security events under temporal change. These studies showed that task adaptation can be moved from parameter updates to instructions, demonstrations, or retrieved context, but the verdict remains part of a generated response.

Recent work has imposed more structure on this inference path. RLFE-IDS employss \textit{Retrieval Augmented Records} (RAG) to retrieve similar traffic through a learned network-specific embedding before asking an LLM to classify the record~\cite{li2025rlfe}. Rehman et al.~\cite{rehman2025flows} combine flow serialization, handcrafted indicators, grammar-constrained JSON, and a decision threshold selected on development data. Their approach is the closest precedent for machine-readable, score-bearing output; however, a constrained generated score is not equivalent to a model interface whose admissible answers are defined before inference.

Studies published in 2026 broaden the evidence on label efficiency and operational cost. Mehavilla et al.~\cite{mehavilla2026evaluating} compare fine-tuned and pretrained language models with conventional machine- and deep-learning baselines, including inference throughput and resource use. Guan et al.~\cite{guan2026notrain} formulate few-shot traffic classification as natural-language meta-tasks without updating model weights. Their measurements show approximately 0.8--4.0s per sample, depending on the LLM, and illustrate that eliminating training does not eliminate per-record inference overhead. Lv and Ding~\cite{lv2026retrieval} likewise combine retrieval, reranking, and few-shot LLM inference without parameter training. LLM-NIRED instead fine-tunes a compact language model for tokenized flow sequences and evaluates transfer to previously unseen networks and deployment on constrained devices~\cite{iannizzotto2026robust}. Uccello and Nadjm-Tehrani~\cite{uccello2026tradeoff} directly examine the trade-off between predictive performance and computational cost for zero- and few-shot LLM inference on 5G telemetry.

\subsection{LLMs as NIDS support components}

Hybrid systems place the LLM outside the per-flow decision path. \emph{Think Fast} runs a lightweight detector at the edge and sends compact telemetry to an LLM for threat interpretation and mitigation advice~\cite{jamshidi2026thinkfast}. SCALA-NIDS assigns the LLM a different control role: it observes runtime statistics and selects a bounded adaptation intent, while a teacher--student detector processes traffic and candidate updates pass through safety validation~\cite{qin2027scala}. These systems are important architectural precursors because they reserve language-model reasoning for higher-level decisions. Their operational measurements, however, characterize alert interpretation or occasional adaptation rather than the cost of using a general model to classify every flow.

\subsection{Positioning of \jevids{}}

The literature therefore covers prompting, retrieval, fine-tuning, constrained generation, cross-network transfer, and LLM-guided adaptation. The Jev applications discussed in Section~\ref{sec:background} concern agentic memory and document coding rather than network traffic. To the best of our knowledge, none of the NIDS studies evaluates a SOM as the direct flow classifier.

\jevids{} addresses this specific question through a controlled comparison of a typed decision model, a structured-output LLM, and a low-data Random Forest. It holds the flow representation and labeled examples constant within each experimental cell, isolates attack names absent from the example pool, and connects predictive outcomes with successful-call latency and list-price cost. Its contribution is this joint evaluation under a matched information budget; it does not claim to introduce prompt-based detection, few-shot classification, or structured output.

\section{Jev-IDS architecture}
\label{sec:architecture}

\begin{figure*}[t]
    \centering
    \includegraphics[width=1.00\linewidth]{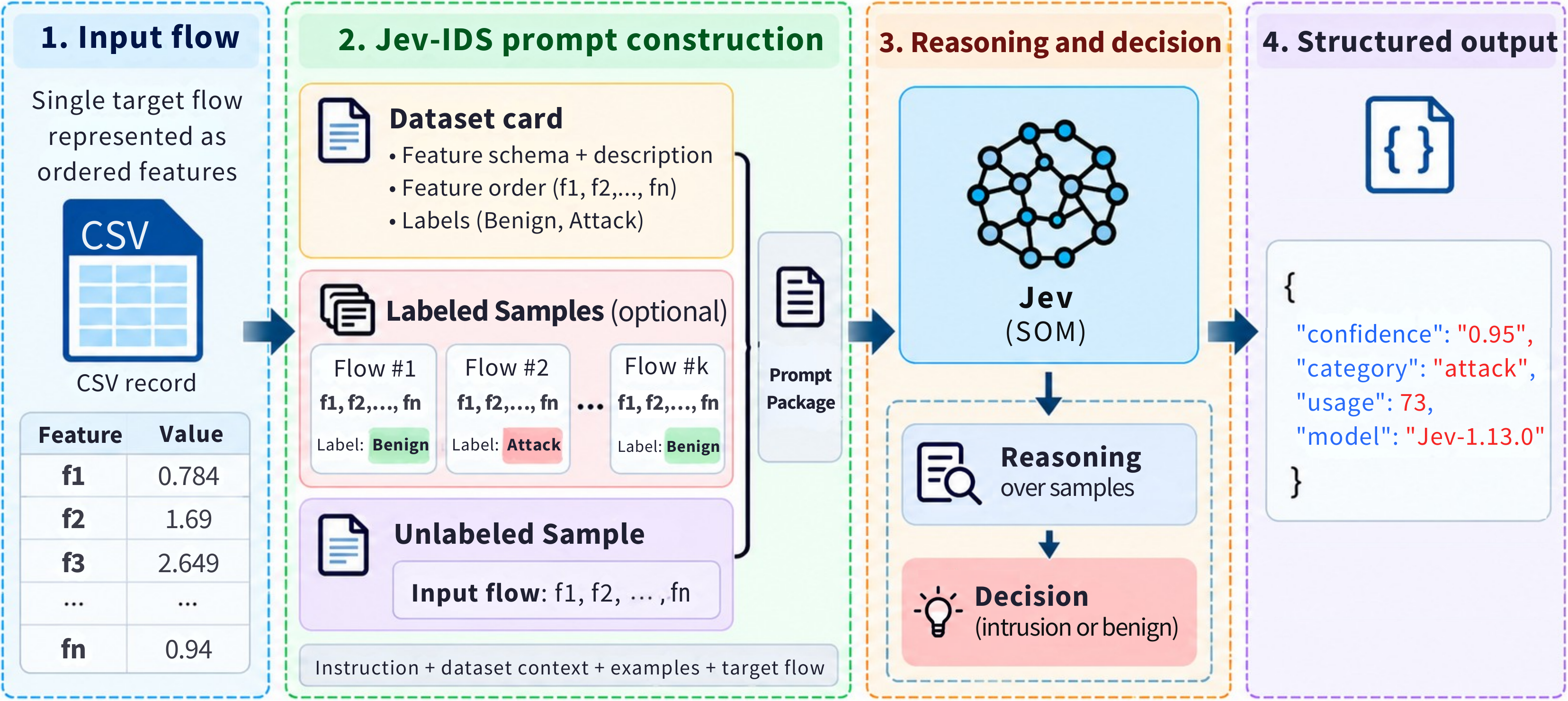}
    \caption{Architecture of \jevids{}. A target Flow is represented by its feature values in Card order and inserted into a \jev{} request together with the request instructions, column names, Category descriptions, and optional labeled Examples. \jev{} answers an intrusion-probability question and a Category question, from which \jevids{} derives the corresponding detection outputs.}
    \label{fig:architecture}
\end{figure*}

\jevids{} is an intrusion-detection pipeline built around \jev{}, a SOM that evaluates typed questions over a supplied state. As illustrated in Figure~\ref{fig:architecture}, \jevids{} processes one network Flow at a time. The pipeline comprises four main stages: the construction of an ordered Flow representation, the construction of the \jev{} request from a fixed request template and optional labeled Examples, the evaluation of two typed questions by \jev{}, and the extraction of the corresponding detection outputs.

\subsection{Flow representation and Dataset Card}

A Dataset is described by a Card that defines its name, ordered features, symbolic features, Categories, and benign Category. A Flow contains the values of these features together with information maintained by \jevids{} about its Category. For the request sent to \jev{}, only the feature values of the target Flow are used. These values are joined as a comma-separated string in the feature order defined by the Card, regardless of the column order in the source CSV file.

The target Flow's Category and attack status are not included in the request sent to \jev{}. Likewise, concrete attack names are not part of the Flow representation supplied to the detector. This distinction is particularly relevant for datasets such as NSL-KDD, in which individual attack names are internally mapped to broader Categories.

\subsection{Jev request construction}

The request submitted to \jev{} is based on a dataset-specific JSON template. The template defines the \jev{} model, a \texttt{state} containing the task instructions, ordered column names, and Category descriptions, and two typed questions. In the current configuration, we explicitly specify the \texttt{jev-1.13.0} \jev{} version. At request time, \jevids{} adds the target Flow to \texttt{state.flows.under\_test}. When labeled Examples are available, they are added to \texttt{state.examples}.

Each Example contains the comma-separated feature values of a Flow and its Category. Examples therefore expose only the Category associated with each reference Flow; concrete attack names are not sent to \jev{}. The Category descriptions already defined in the request state are also used as the criteria of the Category question. The complete request consequently combines the fixed task definition with the target Flow and, when present, the labeled Examples required for that decision.

\subsection{Typed Jev decisions}

Each request asks \jev{} two questions about the same target Flow. The first, identified as \texttt{is\_attack}, is a \texttt{noul} question asking whether the Flow represents an intrusion rather than normal traffic. Its answer is interpreted by \jevids{} as the attack probability, denoted by $p_{\mathrm{attack}}(x)$. The second, identified as \texttt{category}, is a \texttt{choice} question whose alternatives are the Categories defined for the Dataset. Its answer provides the predicted Category and the confidence associated with that choice.

Both questions are evaluated within the same \jev{} request and share the same state. Thus, one request produces the attack probability and Category decision for one target Flow. The \texttt{noul} answer provides the attack probability but does not provide a separate confidence value; the recorded confidence corresponds to the \texttt{choice} answer for the Category. Because these are typed answers, \jevids{} reads the returned fields directly rather than parsing a free-form textual response.

\subsection{Detection output and local verdict}

The answers returned by \jev{} are mapped by the \jevids{} detector interface to three main detection fields: \texttt{p\_attack}, \texttt{category\_pred}, and \texttt{confidence}. Specifically, \texttt{answers.is\_attack.noul} becomes \texttt{p\_attack}, \texttt{answers.category.choice} becomes \texttt{category\_pred}, and \texttt{answers.category.confidence} becomes \texttt{confidence}. The binary Verdict shown in Figure~\ref{fig:architecture} is not returned by \jev{}; it is computed locally from \texttt{p\_attack}. With the current implementation, the decision rule is

\begin{equation}
\hat{y}(x)=
\begin{cases}
1, & p_{\mathrm{attack}}(x)\geq 0.5,\\
0, & p_{\mathrm{attack}}(x)<0.5,
\end{cases}
\label{eq:verdict}
\end{equation}

where $\hat{y}=1$ denotes an attack Verdict and $\hat{y}=0$ denotes a normal Verdict. The resulting detection output therefore separates the probability returned by the \texttt{noul} question, the locally derived binary Verdict, and the Category and confidence returned by the \texttt{choice} question.

The detector interface also retains the model version reported by the response, the request latency, and provider usage information (\textit{i.e.,} numbers of input and output tokens used per request). The complete provider response is retained separately from these extracted detection fields.

\subsection{Request execution and failure handling}

\jevids{} submits one target Flow per \jev{} request. Transient network failures, rate-limit responses, and selected server errors trigger exponential backoff, with up to five total attempts per request. If all applicable attempts fail, the detector returns an error instead of terminating the execution with an exception. This behavior keeps communication failures separate from successful \jev{} responses.

\subsection{Operational placement}

The intended deployment places \jevids{} beside an existing network-monitoring pipeline rather than in the packet-processing path. A sensor or flow exporter can continue to supply Flow information to the existing security infrastructure while copies of Flow records are submitted asynchronously to \jevids{}. In this arrangement, \jevids{} judges one Flow at a time, and its attack probability, Verdict, and Category can be consumed by downstream monitoring or response components.

The intended deployment is therefore an asynchronous side channel rather than an inline packet filter. A queue can decouple the flow-export process from \jevids{} processing. When the hosted \jev{} service is used, the selected Flow features are transmitted to the remote service, while packet payloads are not part of the request.

\section{Experimental methodology}
\label{sec:method}

The experimental methodology was designed to assess \jevids{} under different levels of labeled-data availability and to compare its behavior with representative generative, supervised, and unsupervised approaches. All detectors are evaluated on the same target flows and feature representation whenever applicable, while the number of labeled examples available during prediction or training is explicitly controlled. This design allows us to examine not only overall detection performance, but also the ability of each approach to recognize attacks that are not represented among the labeled examples.

\subsection{Experimental design}

The experiments consider three factors: the detector, the number of labeled examples per category ($k$), and the sampling seed. For each seed, labeled examples are sampled from the training pool independently of the evaluation set. Sampling is nested across values of $k$, such that examples available under a smaller budget remain available when the budget is increased. The same sampled examples are used by all detectors evaluated under the corresponding condition.

For \jev{} and the generative baseline, we consider
$k\in\{0,1,2,4,8\}$. The $k=0$ condition corresponds to zero-shot inference, whereas $k>0$ provides $k$ labeled flows from each of the five traffic categories. The Random Forest is evaluated with $k\in\{1,2,4,8\}$ and additionally with the complete labeled training pool, providing a conventional fully supervised reference. Three independent sampling seeds are used for the finite-$k$ conditions.

Each target flow is classified independently. For detectors producing an attack probability, the binary decision is obtained using a common decision threshold. Outputs that do not provide a valid prediction are accounted for separately through the call-error rate and are treated as benign when computing the binary detection metrics, following a fail-open convention.

\subsection{Comparison methods}

\paragraph{Structured-output LLM.}
Gemini~3.6~Flash is used as the generative baseline. It receives the same ordered flow representation, traffic-category descriptions, and labeled examples supplied to \jev{} under the corresponding experimental condition. The model produces a structured output containing an attack probability, a binary classification, and a predicted traffic category. Deterministic generation settings are used to reduce variability unrelated to the labeled-example sampling process.

\paragraph{Random Forest.}
A Random Forest with 100 trees is used as the supervised machine-learning baseline. Numerical attributes are represented directly, whereas categorical attributes are one-hot encoded. For each finite value of $k$, the classifier is trained using exactly the same labeled flows made available to \jev{} and the generative baseline. An additional condition trains the classifier using the complete labeled training pool. The attack probability is obtained from the class-probability estimates by aggregating the prediction into benign and attack classes.

\paragraph{Isolation Forest.}
An Isolation Forest with 100 trees provides an unsupervised reference. Unlike the other approaches, it does not use labeled attack examples and is trained exclusively on benign flows from the training pool. Its anomaly score is subsequently used to distinguish normal from anomalous traffic. This baseline represents a detection setting in which no attack examples are available during training.

\subsection{Dataset and evaluation protocol}

The experiments use the NSL-KDD dataset. Although NSL-KDD does not reflect contemporary network traffic, it remains one of the most widely used historical benchmarks in intrusion-detection research, enabling direct comparison with a broad body of prior work. Moreover, its well-defined separation between training and test sets allows attack types present and absent from the labeled pool to be explicitly distinguished, which is particularly suitable for studying detector behavior under limited labeled-data availability. We therefore use NSL-KDD as a methodological benchmark for controlled comparison and reproducibility, rather than as a proxy for current operational networks; validation on more recent datasets is considered complementary to this analysis.

KDDTrain+ constitutes the training pool from which labeled examples are obtained, while evaluation flows are sampled exclusively from KDDTest+. Consequently, no evaluation flow is used for selecting the examples supplied to the detectors or for fitting the supervised and unsupervised baselines. The evaluation set contains 2,000 flows: 874 normal flows and 1,126 attacks. The attack instances comprise 642 denial-of-service (DoS), 248 remote-to-local (R2L), 213 probe, and 23 user-to-root (U2R) flows. The set is fixed before detector execution and shared across all experimental conditions.

\begin{table}[t]
\centering
\caption{Composition of the NSL-KDD evaluation set.}
\label{tab:datasets}
\small
\resizebox{\columnwidth}{!}{%
\begin{tabular}{lrrrrr}
\toprule
Flows & Normal & DoS & Probe & R2L & U2R \\
\midrule
2,000 & 874 & 642 & 213 & 248 & 23 \\
\bottomrule
\end{tabular}
}
\end{table}

To evaluate generalization beyond the attack types represented in the training pool, attack instances are further divided into \emph{known} and \emph{novel} attacks. An attack is considered novel when its specific attack type occurs in KDDTest+ but is absent from KDDTrain+. Under this definition, 300 of the 1,126 attack flows are novel, whereas the remaining 826 belong to attack types already represented in the training pool. Specific attack names are not provided to the evaluated detectors; their outputs are restricted to the five coarse traffic categories: normal, DoS, probe, R2L, and U2R.

\subsection{Evaluation metrics}

The primary detection metric is the attack-class F1 score,

\begin{equation}
F_1 = \frac{2TP}{2TP+FP+FN}.
\label{eq:f1}
\end{equation}

We additionally report precision, attack recall, PR-AUC, ROC-AUC, and recall for each traffic category. To examine generalization to previously unseen attack types, recall is also computed separately for known and novel attacks. For subset-specific F1 analyses, the selected attack subset is evaluated together with the complete set of normal flows, ensuring that false positives remain represented in the metric.

Metrics for the finite-$k$ conditions are first computed independently for each sampling seed and subsequently summarized across the three seeds. F1 is reported using the mean and sample standard deviation.

In addition to predictive performance, we measure inference latency and, for externally hosted models, token consumption and estimated inference cost. Cost estimates are normalized per one million flow classifications and are calculated from the public list prices applicable at the time of the experiments. These measurements are intended to characterize the operational trade-offs among the evaluated approaches rather than only their predictive accuracy.

\subsection{Statistical analysis}

Pairwise comparisons between detectors are performed on predictions obtained for the same evaluation flows under equivalent values of $k$ and sampling seeds. Differences in binary classification outcomes are assessed using the exact two-sided McNemar test \cite{mcnemar1947note}. The analysis considers the discordant predictions between each pair of detectors, thereby testing whether their error patterns differ significantly on the same observations.

Comparisons involving different supervision regimes, such as a few-shot detector and a Random Forest trained using the complete labeled pool, are treated separately and explicitly identified as cross-budget comparisons.

\subsection{Reproducibility}

The evaluation set, labeled-example selections, detector configurations, prompts, and source-code revision are fixed and recorded for each experiment. Prediction-level outputs are retained together with the corresponding experimental condition, enabling the reported aggregate metrics and paired statistical comparisons to be reproduced from the original decisions. The implementation and experimental artifacts are publicly available at
\url{https://github.com/jev-ids/jev-ids}.

\section{Results}
\label{sec:results}

This section evaluates \jevids{} from three complementary perspectives. First, we examine predictive performance as the number of labeled Examples increases and compare it with the generative, supervised, and unsupervised baselines. We then investigate generalization to attack types absent from the labeled pool and analyze whether the observed differences are statistically significant. Finally, we assess the operational trade-offs of the hosted model-based Detectors in terms of latency, token usage, and estimated inference cost. Unless otherwise stated, results for finite values of $k$ are averaged over three independently sampled Example sets evaluated on the same 2,000 target Flows.

\subsection{Overall detection performance}

Table~\ref{tab:paper_results} summarizes the predictive results on the NSL-KDD evaluation set. Each finite-$k$ condition comprises 6,000 decisions, corresponding to the same 2,000 target Flows evaluated under three independently sampled Example sets.

\begin{table*}[t]
\centering
\caption{Predictive results on the 2,000-Flow NSL-KDD evaluation set. F1 is reported as mean $\pm$ sample standard deviation across three Example seeds. The remaining metrics are seed means.}
\label{tab:paper_results}
\scriptsize
\resizebox{\textwidth}{!}{%
\begin{tabular}{llrrrrrrr}
\toprule
Detector & $k$ & F1 & Precision & Recall & Known recall & Novel recall & PR-AUC & ROC-AUC \\
\midrule
\multirow{5}{*}{\jev{}}
 & 0 & $0.782\pm0.005$ & 0.972 & 0.654 & 0.665 & 0.622 & 0.940 & 0.925 \\
 & 1 & $0.856\pm0.025$ & 0.953 & 0.778 & 0.790 & 0.747 & 0.952 & 0.950 \\
 & 2 & $0.846\pm0.009$ & 0.949 & 0.763 & 0.772 & 0.738 & 0.949 & 0.944 \\
 & 4 & $0.856\pm0.015$ & 0.944 & 0.784 & 0.803 & 0.731 & 0.959 & 0.959 \\
 & 8 & $0.854\pm0.017$ & 0.942 & 0.783 & 0.805 & 0.721 & 0.964 & 0.962 \\
\midrule
\multirow{5}{*}{Gemini 3.6 Flash}
 & 0 & $0.867\pm0.005$ & 0.946 & 0.801 & 0.842 & 0.687 & 0.949 & 0.959 \\
 & 1 & $0.880\pm0.020$ & 0.942 & 0.826 & 0.866 & 0.713 & 0.948 & 0.959 \\
 & 2 & $0.869\pm0.017$ & 0.943 & 0.806 & 0.849 & 0.684 & 0.947 & 0.959 \\
 & 4 & $0.884\pm0.002$ & 0.942 & 0.833 & 0.875 & 0.717 & 0.946 & 0.958 \\
 & 8 & $0.881\pm0.012$ & 0.949 & 0.823 & 0.867 & 0.702 & 0.951 & 0.961 \\
\midrule
\multirow{5}{*}{Random Forest}
 & 1 & $0.748\pm0.036$ & 0.598 & 1.000 & 1.000 & 1.000 & 0.791 & 0.740 \\
 & 2 & $0.777\pm0.062$ & 0.644 & 0.987 & 0.988 & 0.983 & 0.875 & 0.854 \\
 & 4 & $0.853\pm0.011$ & 0.771 & 0.955 & 0.969 & 0.914 & 0.913 & 0.907 \\
 & 8 & $0.865\pm0.013$ & 0.794 & 0.950 & 0.964 & 0.912 & 0.935 & 0.922 \\
 & all & $0.765\pm0.000$ & 0.971 & 0.631 & 0.764 & 0.263 & 0.963 & 0.964 \\
\midrule
Isolation Forest
 & all & $0.761\pm0.000$ & 0.974 & 0.624 & 0.643 & 0.573 & 0.958 & 0.945 \\
\bottomrule
\end{tabular}
}
\end{table*}

Gemini achieved higher overall F1 than \jev{} at every evaluated value of $k$. At $k=1$, for example, Gemini reached an F1 of 0.880, compared with 0.856 for \jev{}. This difference was primarily associated with recall: Gemini detected 82.6\% of the attacks, whereas \jev{} detected 77.8\%, while both maintained similar precision (0.942 and 0.953, respectively). For \jev{}, most of the improvement over the zero-shot condition occurred with the introduction of a single Example per Category, increasing F1 from 0.782 to 0.856. Additional Examples produced only small variations, with F1 remaining between 0.846 and 0.856 for $k\geq1$. A similarly non-monotonic behavior was observed for Gemini.

The few-shot Random Forest exhibited a markedly different precision--recall trade-off. At $k=1$, it detected all attacks but obtained only 0.598 precision, corresponding to approximately 764 false alarms among the 874 normal Flows in each seed, compared with approximately 43 for \jev{}. Increasing the labeled budget progressively reduced this behavior: at $k=8$, the Random Forest reached an F1 of 0.865 and a precision of 0.794, although it still produced substantially more false alarms than either model-based Detector. Interestingly, training the Random Forest on the complete labeled pool did not improve its performance on the evaluation set. Its precision increased to 0.971, but recall decreased to 0.631 and novel-attack recall to 0.263.

The Isolation Forest showed a similarly conservative operating profile, reaching 0.974 precision and 0.624 recall. Its novel-attack recall of 0.573 was higher than that of the fully supervised Random Forest, but below the corresponding few-shot \jev{} and Gemini conditions. Taken together, these results show that overall F1 alone does not fully characterize the differences among the Detectors, particularly when the cost of false alarms and the availability of labeled examples are considered.

The threshold-independent metrics provide an additional perspective. At $k=1$, \jev{} reached PR-AUC~=~0.952 and ROC-AUC~=~0.950, compared with 0.948 and 0.959 for Gemini. At $k=8$, the corresponding values were 0.964 and 0.962 for \jev{}, and 0.951 and 0.961 for Gemini. The fully supervised Random Forest also achieved high PR-AUC and ROC-AUC (0.963 and 0.964) despite an F1 of only 0.765 at the common decision threshold. This discrepancy indicates that part of the observed performance difference is associated with the operating point rather than with the ranking of attack scores alone. Since probability calibration was not evaluated, these scores are not interpreted as calibrated probabilities.

\subsection{Generalization to novel attacks}

A central question in our evaluation is whether labeled Examples help the Detectors recognize attack types that are not themselves represented in the labeled pool. Without Examples ($k=0$), Gemini obtained higher recall on novel attacks than \jev{} (0.687 versus 0.622). This relationship changed once labeled Examples were introduced. \jev{} reached novel-attack recalls of 0.747, 0.738, 0.731, and 0.721 for $k=1,2,4,8$, respectively, compared with 0.713, 0.684, 0.717, and 0.702 for Gemini.

The difference is not uniform across attack Categories. The largest separation occurs for novel R2L attacks. With labeled Examples, \jev{} obtained recalls of 0.151, 0.198, 0.262, and 0.222 for $k=1,2,4,8$, respectively, whereas Gemini obtained 0.071, 0.000, 0.008, and 0.008. Conversely, Gemini remained stronger for several other attack types, most notably probe traffic. The aggregate novel-attack results therefore reflect different generalization behavior across attack families rather than a consistent advantage for one Detector on every unseen attack type.

The Random Forest illustrates why novel-attack recall should not be considered in isolation. In the few-shot conditions it reached recall between 0.912 and 1.000 on novel attacks, but at the cost of substantially lower precision. At $k=1$, for instance, its perfect novel-attack recall coincided with a precision of only 0.598 over the complete evaluation set. Thus, the ability to detect novel attacks must be considered together with the corresponding false-alarm rate.

\subsection{Paired statistical comparison}

Table~\ref{tab:mcnemar} reports the paired comparison between \jev{} and Gemini using the exact McNemar test. Each all-Flow comparison contains 6,000 paired decisions, while the known- and novel-attack subsets contain 2,478 and 900 decisions, respectively.

\begin{table*}[t]
\centering
\caption{Paired comparison between \jev{} and Gemini~3.6~Flash. The ``Jev correct'' and ``Gemini correct'' columns report the correctly classified instances among discordant decisions.}
\label{tab:mcnemar}
\small
\resizebox{0.75\textwidth}{!}{%
\begin{tabular}{lrrrrrr}
\toprule
Subset & $k$ & Pairs & Discordant & Jev correct & Gemini correct & Exact $p$ \\
\midrule
All Flows & 0 & 6,000 & 833 & 213 & 620 & $<0.001$ \\
All Flows & 1 & 6,000 & 508 & 195 & 313 & $<0.001$ \\
All Flows & 2 & 6,000 & 515 & 199 & 316 & $<0.001$ \\
All Flows & 4 & 6,000 & 506 & 178 & 328 & $<0.001$ \\
All Flows & 8 & 6,000 & 472 & 159 & 313 & $<0.001$ \\
\midrule
Known attacks & 0 & 2,478 & 489 & 25 & 464 & $<0.001$ \\
Known attacks & 1 & 2,478 & 266 & 38 & 228 & $<0.001$ \\
Known attacks & 2 & 2,478 & 235 & 22 & 213 & $<0.001$ \\
Known attacks & 4 & 2,478 & 227 & 24 & 203 & $<0.001$ \\
Known attacks & 8 & 2,478 & 190 & 18 & 172 & $<0.001$ \\
\midrule
Novel attacks & 0 & 900 & 206 & 74 & 132 & $<0.001$ \\
Novel attacks & 1 & 900 & 118 & 74 & 44 & 0.007 \\
Novel attacks & 2 & 900 & 162 & 105 & 57 & $<0.001$ \\
Novel attacks & 4 & 900 & 167 & 90 & 77 & 0.353 \\
Novel attacks & 8 & 900 & 171 & 94 & 77 & 0.221 \\
\bottomrule
\end{tabular}
}
\end{table*}

For the complete evaluation set, Gemini was correct on more discordant decisions at every value of $k$, with $p<0.001$ in all comparisons. The difference was particularly pronounced for known attack types, for which Gemini was also favored at every labeled-example budget. These results are consistent with the higher overall and known-attack recall reported in Table~\ref{tab:paper_results}.

A different pattern emerges for novel attacks. At $k=1$, \jev{} was correct on 74 of the 118 discordant decisions, compared with 44 for Gemini ($p=0.007$). The difference increased at $k=2$, where \jev{} was correct on 105 discordant decisions and Gemini on 57 ($p<0.001$). Although \jev{} retained numerically higher novel-attack recall at $k=4$ and $k=8$, the paired tests did not identify a statistically significant difference at these budgets ($p=0.353$ and $p=0.221$, respectively).

The paired analysis against the low-data Random Forest further illustrates the effect of limited supervision. At $k=1$, \jev{} and the Random Forest disagreed on 2,909 of the 6,000 paired decisions; \jev{} was correct in 2,161 of these cases, yielding a significant difference ($p<0.001$). As more labeled data became available, the gap narrowed, and at $k=8$ the Random Forest reached a slightly higher F1 than \jev{} (0.865 versus 0.854). This behavior is consistent with the Random Forest progressively reducing the large number of false positives observed under the smallest labeled-data budgets.

\subsection{Operational efficiency}

Beyond predictive performance, an API-based IDS must also account for the computational cost of evaluating each Flow. Table~\ref{tab:operational} reports the observed request latency and estimated list-price cost for \jev{} and Gemini. Across the evaluated Example budgets, \jev{} required approximately 0.31--0.34~s per successful request, compared with 1.99--2.65~s for Gemini. Its estimated cost ranged from US\$43 to US\$295 per million Flow decisions, compared with US\$1,068 to US\$3,035 for Gemini.

\begin{table*}[t]
\centering
\caption{Operational comparison between \jev{} and Gemini~3.6~Flash. Cost denotes the estimated list price per one million Flow decisions, and latency denotes the mean wall-clock duration of a successful model request. Ratios are Gemini/\jev{}.}
\label{tab:operational}
\small
\resizebox{0.8\textwidth}{!}{%
\begin{tabular}{rrrrrrr}
\toprule
$k$ & \jev{} cost & Gemini cost & Cost ratio & \jev{} latency & Gemini latency & Latency ratio \\
\midrule
0 & US\$43  & US\$1,068 & $25\times$ & 310 ms & 2.24 s & $7.2\times$ \\
1 & US\$74  & US\$1,651 & $22\times$ & 315 ms & 2.42 s & $7.7\times$ \\
2 & US\$106 & US\$2,409 & $23\times$ & 308 ms & 2.65 s & $8.6\times$ \\
4 & US\$169 & US\$2,869 & $17\times$ & 322 ms & 2.13 s & $6.6\times$ \\
8 & US\$295 & US\$3,035 & $10\times$ & 335 ms & 1.99 s & $5.9\times$ \\
\bottomrule
\end{tabular}
}
\end{table*}

At $k=1$, where each Detector receives five labeled Examples in total, \jev{} exhibited approximately $7.7\times$ lower request latency and a $22\times$ lower estimated list-price cost than Gemini. This operational difference accompanied an F1 reduction of 0.024, while \jev{} maintained slightly higher precision (0.953 versus 0.942). Across all evaluated budgets, the latency ratio ranged from $5.9\times$ to $8.6\times$, whereas the estimated cost ratio ranged from $10\times$ to $25\times$.

The cost behavior also reflects differences in token usage and provider accounting. At $k=1$, a \jev{} decision uses approximately 1.8k input tokens, with no output-token charge under the pricing model used in the experiments. Gemini additionally incurs output charges, including approximately 200 reasoning tokens per Flow under its lowest reasoning setting. As the number of Examples increases, prompt length becomes a larger component of the inference cost. At $k=8$, Gemini reported an average of 6,269 input tokens per request, of which 4,004 were served from provider-side cache. This caching contributed to the reduction in the cost ratio observed at the largest Example budget.

These values characterize the hosted services under the experimental conditions and pricing in effect at the time of measurement. In particular, latency includes network and provider-side overhead and should therefore be interpreted as observed end-to-end request latency rather than intrinsic model execution time. Similarly, the reported costs are normalized list-price estimates derived from the recorded token usage and are intended to characterize the relative operational requirements of the two approaches rather than future or deployment-specific billing.

\section{Discussion}
\label{sec:discussion}

The results position \jev{} differently from both the generative LLM and the conventional machine-learning baselines. Gemini achieves higher overall F1 across the evaluated labeled-example budgets, largely through higher attack recall, whereas \jev{} maintains comparable precision with substantially lower inference latency and estimated cost. Thus, the main finding is not that a SOM consistently surpasses a generative LLM in detection accuracy, but that it provides a distinct trade-off between predictive performance and inference efficiency. This distinction is relevant for IDS scenarios in which model invocation cost and response time constrain how broadly an AI-based detector can be applied.

A second finding concerns generalization beyond the attack types represented in the labeled pool. Once labeled Examples are introduced, \jev{} consistently obtains higher recall than Gemini on novel attack types, although the statistical evidence is strongest at the smallest Example budgets. This effect should not be interpreted as a general superiority in detecting unseen attacks. 

The Category-level analysis shows that a substantial part of the difference is associated with R2L traffic, while Gemini remains stronger for other attack families. More broadly, neither model benefits monotonically from additional Examples. For \jev{}, most of the improvement over zero-shot inference occurs with only a small amount of labeled evidence, after which performance reaches a relatively stable region. This suggests that, in the present setting, the value of additional Examples depends more on their informational content than on their quantity alone.

The behavior of the conventional baselines reinforces the importance of considering the complete error profile rather than a single aggregate metric. Under very limited supervision, the Random Forest achieves high attack recall by classifying a large fraction of benign traffic as malicious. Additional labeled data progressively reduces this tendency, and its overall F1 eventually becomes comparable to that of the model-based Detectors, although with substantially lower precision. Conversely, training the Random Forest on the complete labeled pool yields high precision but considerably lower recall, particularly for novel attack types. The Isolation Forest exhibits a similarly conservative profile. These contrasting behaviors illustrate that comparable F1 values may correspond to very different operational consequences for an IDS, especially when false alarms impose a significant burden on downstream analysis.

The threshold-independent results provide additional context for these differences. In particular, the fully supervised Random Forest exhibits strong PR-AUC and ROC-AUC despite considerably lower F1 at the common decision threshold. This indicates that part of the observed difference among Detectors stems from the selected operating point rather than exclusively from their ability to rank malicious and benign Flows. 

Consequently, the use of a common threshold facilitates controlled comparison but should not be interpreted as an optimized deployment configuration for every Detector. Threshold selection and probability calibration therefore remain important dimensions for future evaluation.

Operational efficiency constitutes another relevant distinction between the two hosted model-based approaches. Under the evaluated conditions, \jev{} requires substantially less wall-clock time and has a markedly lower estimated list-price cost than Gemini. The difference remains across all labeled-example budgets, although the cost gap narrows as prompt size increases and provider-side caching becomes more relevant. These observations suggest that SOM-based inference may be attractive in settings where a model must be invoked repeatedly over large numbers of network Flows. At the same time, both hosted approaches remain considerably slower than local machine-learning inference. The present results therefore favor their use in roles such as alert enrichment, secondary analysis, or selective escalation rather than as evidence of suitability for line-rate packet processing.

This observation also motivates hybrid detection strategies. Because \jev{} exposes a structured attack score together with its predicted Category, it could be incorporated into a cascade in which straightforward cases are handled by a lower-cost component and uncertain cases are escalated to a more expensive generative model or to a human analyst. Such an architecture could combine the efficiency observed for \jev{} with the stronger overall recall of the LLM. This possibility is not evaluated in the present study and should therefore be regarded as a direction for future investigation rather than as an established benefit of \jevids{}.

The findings should also be interpreted in light of the experimental scope. NSL-KDD provides a controlled and widely used benchmark that supports reproducible comparison with prior IDS research, but it does not represent contemporary operational traffic. The definition of novelty used here is likewise specific: an attack is considered novel when its concrete attack type is absent from the training pool, which does not necessarily imply a fundamentally new behavior or vulnerability. Consequently, the observed differences in novel-attack recall should be understood as evidence of generalization across withheld attack types within this benchmark rather than as direct evidence of zero-day detection in operational networks.

Several extensions are therefore necessary before broader conclusions can be drawn. Evaluation on more recent and heterogeneous datasets is the most immediate step, together with experiments based on temporal or organizational separation between training and test traffic. Future work should also examine probability calibration, detector-specific threshold selection on an independent validation set, and stronger tabular-learning baselines. 

Finally, statistical analyses should account explicitly for the repeated evaluation of the same target Flows across different Example draws, while additional seeds would provide a more robust characterization of the variability introduced by few-shot sampling.

\section{Conclusion}
\label{sec:conclusion}

This work investigated whether a SOM can be effectively used for flow-based network intrusion detection with limited labeled supervision. We proposed \jevids{}, a detection pipeline built around \jev{} that combines an ordered Flow representation, optional Category-labeled Examples, typed attack and Category questions, and structured outputs. The proposal differs from conventional supervised approaches by avoiding task-specific model training while still providing a simple pipeline with traceable inference records.

The results indicate that \jev{} can achieve strong detection performance even when only a few labeled Examples are available. In the NSL-KDD pilot, it obtained the best overall results among the evaluated approaches for the considered settings, with fewer false alarms than the Random Forest and better detection of novel attacks than the GPT-5.6 Luna. The results also showed favorable latency and estimated inference cost in the setting with the fewest labeled Examples. 

Future work will extend the experimental conditions, including additional values of $k$ and evaluation on additional datasets. We also intend to strengthen the statistical analysis by accounting for the dependence introduced by seeds and repetitions, and to further investigate probability calibration, robustness to novel attacks, and operational behavior under realistic network conditions. 

\section*{Declaration of competing interest}

The authors declare that they have no known competing financial interests or personal relationships that could have appeared to influence the work reported in this paper. 

\section*{Funding}

This work was supported by the Fundação de Amparo à Pesquisa do Estado do Rio Grande do Sul (FAPERGS), Brazil, under Grant 26/2551-9000194-3.

\section*{Data and code availability}

NSL-KDD is a third-party dataset and should be obtained from their original distribution points under their respective terms. The MIT-licensed experiment code, dataset preparation scripts, frozen split identifiers, prompts, raw pilot prediction rows, and result summaries are public at \url{https://github.com/jev-ids/jev-ids} \cite{jevids2026software}; project documentation is available at \url{https://jev-ids.github.io/}. The revision analyzed here corresponds to commit \texttt{286a310}.%; the full identifier is recorded in the software reference. \todo{Create and cite an immutable archival release/DOI for the submission artifact.} Raw hosted-model response bodies should be reviewed for redistribution constraints before release.

\bibliographystyle{abbrv}
\bibliography{cas-refs}

\end{document}